\documentclass[numberedappendix,iop]{emulateapj}

\usepackage{graphicx}
\usepackage{natbib}
\usepackage{ulem}
\usepackage{txfonts}

\slugcomment{Submitted to the Astrophysical Journal}

\shorttitle{}
\shortauthors{Mart\' inez Gonz\'alez et al.}

\begin{document}

\title{Resolving the internal magnetic structure of the solar network}
\author{M. J. Mart\'{\i}nez Gonz\'alez}
\affil{Instituto de Astrof\' isica de Canarias, V\' ia L\'actea s/n, E-38205 La
 Laguna, Tenerife, 
Spain}
\affil{Departamento de Astrof\'\i sica, Universidad de La Laguna, E-38205 La Laguna, Tenerife, Spain}

\author{L. R. Bellot Rubio}
\affil{Instituto de Astrof\' isica de Andaluc\' ia (CSIC), Apdo. de Correos 3004, 18080 Granada, Spain}

\author{S. K. Solanki}
\affil{Max-Planck-Institut f\"ur Sonnensystemforschung, 37191, Katlenburg-Lindau, Germany\\
School of Space Research, Kyung Hee University, Yongin, Gyeonggi, 446-701 Korea}

\author{V. Mart\' inez Pillet} 
\affil{Instituto de Astrof\' isica de Canarias, V\' ia L\'actea s/n, E-38205 La Laguna, Tenerife, 
Spain}
\affil{Departamento de Astrof\'\i sica, Universidad de La Laguna, E-38205 La Laguna, Tenerife, Spain}

\author{J. C. Del Toro Iniesta}
\affil{Instituto de Astrof\' isica de Andaluc\' ia (CSIC), Apdo. de Correos 3004, 18080 Granada, Spain}

\author{P. Barthol} 
\affil{Max-Planck-Institut f\"ur Sonnensystemforschung, 37191, Katlenburg-Lindau, Germany}

\author{W. Schmidt}
\affil{KIS Kiepenheuer-Institut f\"ur Sonnenphysik, D-79104 Freiburg, Germany}

\begin{abstract}

We analyze the spectral asymmetry of Stokes $V$ (circularly polarized) profiles of an individual network patch in the quiet Sun observed by Sunrise/IMaX. At a spatial resolution of $0.15''-0.18''$, the network elements contain substructure which is revealed by the spatial distribution of Stokes $V$ asymmetries. The area asymmetry between the red and blue lobes of Stokes $V$ increases from nearly zero at the core of the structure to values close to unity at its edges (one-lobed profiles). Such a distribution of the area asymmetry is consistent with magnetic fields expanding with height, i.e., an expanding magnetic canopy (which is required to fulfill pressure balance and flux conservation in the solar atmosphere). Inversion of the Stokes $I$ and $V$ profiles of the patch confirms this picture, revealing a decreasing field strength and increasing height of the canopy base from the core to the periphery of the network patch. However, the non-roundish shape of the structure and the presence of negative area and amplitude asymmetries reveal that the scenario is more complex than a canonical flux tube expanding with height surrounded by downflows.
\end{abstract}


\section{Introduction}

Asymmetries in spectral profiles of Stokes $V$ are commonly measured on the solar surface, from the penumbrae of sunspots to the quiet Sun; they have been known for decades \citep[e.g.][]{sami_84, grossmann_96, valentin_97, sigwarth_99, sigwarth_01, rezaei_steiner_07, marian_08, bartolomeo_11}. The generally accepted model to produce asymmetric Stokes $V$ profiles is by having anticorrelated velocity and magnetic field gradients along the line of sight \citep{illing_75, jorge_89, arturo_02}. Therefore, circular polarization asymmetries encode the structuring of magnetic fields along the line of sight. However, the limited spatial resolution of most observations mixes the contributions of different regions to the asymmetries, hence, allowing only indirect inferences of gradients along the line of sight.

Outside of sunspots the classical model to produce asymmetries is at the canopies of the magnetic feaures expanding with height \citep{grossmann_88, sami_89}. This model has been able to reproduce the simultaneous 
presence of Stokes $V$ area asymmetry, the absence of a Stokes $V$ zero-crossing shift \citep{sami_86, valentin_97}, and the change in sign of the area asymmetry near the solar limb \citep{bunte_93} in obervations of the quiet Sun and active region plage taken at rather low spatial resolution \citep{stenflo_solanki_87, valentin_97}. Alternatives have also been suggested, such as the presence of bulk velocities in the flux tube \citep{luis_97, luis_00,sigwarth_01}, or the MISMA and MESMA models \citep{jorge_96, thorsten_markus_07}. 

This letter is devoted to employing Stokes $V$ asymmetry in order to diagnose the internal structure of network magnetic features, i.e., the stongest flux accumulations in the quiet Sun. It is known, since the work of \cite{stenflo_73}, that network patches are composed of magnetic elements with kG fields. Later, two very different models have been proposed to explain the observed asymmetries. First, the thin flux tube model that assumed a rather vertical magnetic tube expanding with height. Second, the MISMA hypothesis that assumes an infinite number of magnetic atmospheres in the resolution element. In both cases, the cause for the asymmetries is the correlation between the distribution of magnetic fields and velocities. However, the MISMA model is a one-dimensional model, having no predictive power for the spatial variation of asymmetries in magnetic structures. In order to understand how magnetic fields are really organized and to discern between these two models (or create new ones), we must observe at higher spatial resolutions, which will allow us to directly map the substructure of magnetic features and to trace gradients along the optical depth.

\section{Data and analysis}

We map the Stokes $V$ profile asymmetries in the quiet Sun using data obtained by IMaX \citep{valentin_11} on Sunrise \citep{sami_10, barthol_11}. The data analyzed here consist of Stokes $I$ and $V$ profiles of the Fe\,{\sc i} 5250.2 \AA\ line (Land\'e factor of 3) sampled at 11 wavelegth positions that were recorded on June 10th at the disk center. The effective field of view (FOV) is 46.8$''\times$46.8$''$  and the spatial resolution is about 0.2$''$. The noise level in Stokes $V$ is $10^{-3}$ in units of the continuum intensity, $I_c$, prior to the application 
of the phase diversity reconstruction technique. After this procedure, the noise of the reconstructed data is spatially and temporaly correlated, but still follows a Gaussian distribution, with a standard deviation of $2.5 \times 10^{-3}$ $I_c$. 

For the analysis presented in this letter, we select a small region of the FOV containing the largest and the most intense circular polarization patch (see Fig. \ref{mapas_region}). It has a magnetic flux of $\sim 2\times 10^{19}$ Mx contained in an area of $\sim 5.5$ Mm$^2$. The pattern of the granulation is distorted within the magnetic patch, which is typical of kG field accumulations. In order to minimize the effect of the noise in our analysis, we select only those profiles with amplitudes larger than 5 times the noise level. 46.6\% of the non-reconstructed, and 29.3\% of the reconstructed profiles belonging to the small selected area meet this criterion. The selected Stokes $V$ profiles in this region are mostly regular profiles. More precisely, 93.6\% and 88.9\% of the profiles have two lobes in the non-reconstructed and in the phase diversity reconstructed data, respectively. We define the area ($\delta A$) and amplitude ($\delta a$) asymmetries of regular profiles as:

\begin{figure}[!t]
\centering{
\includegraphics[width=0.25\textwidth, bb=173 434 500 686]{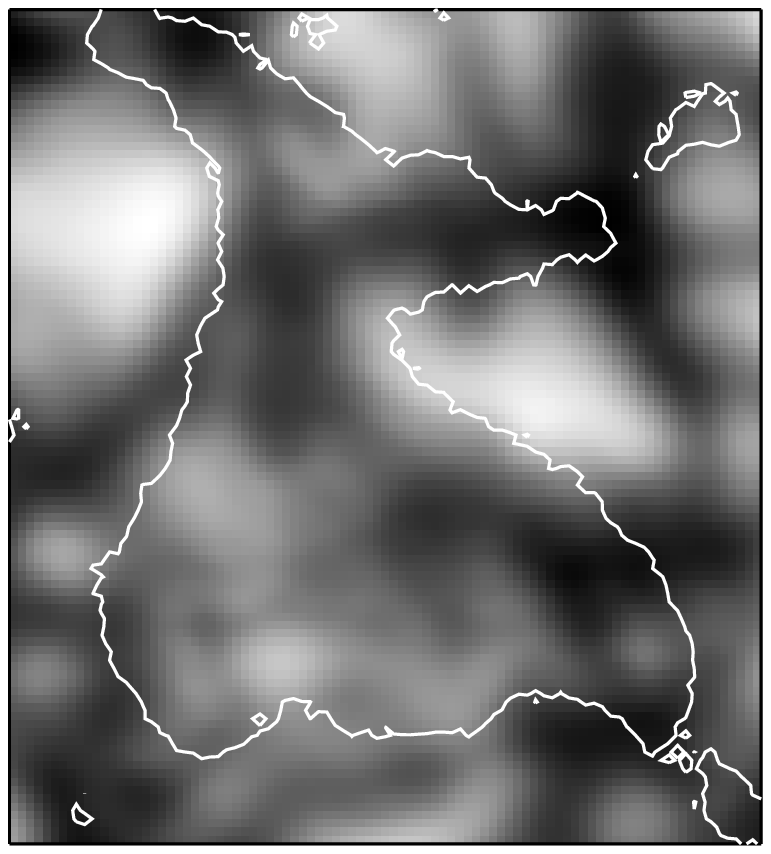}
\hspace{-1.5cm}
\includegraphics[width=0.25\textwidth, bb=173 434 500 686]{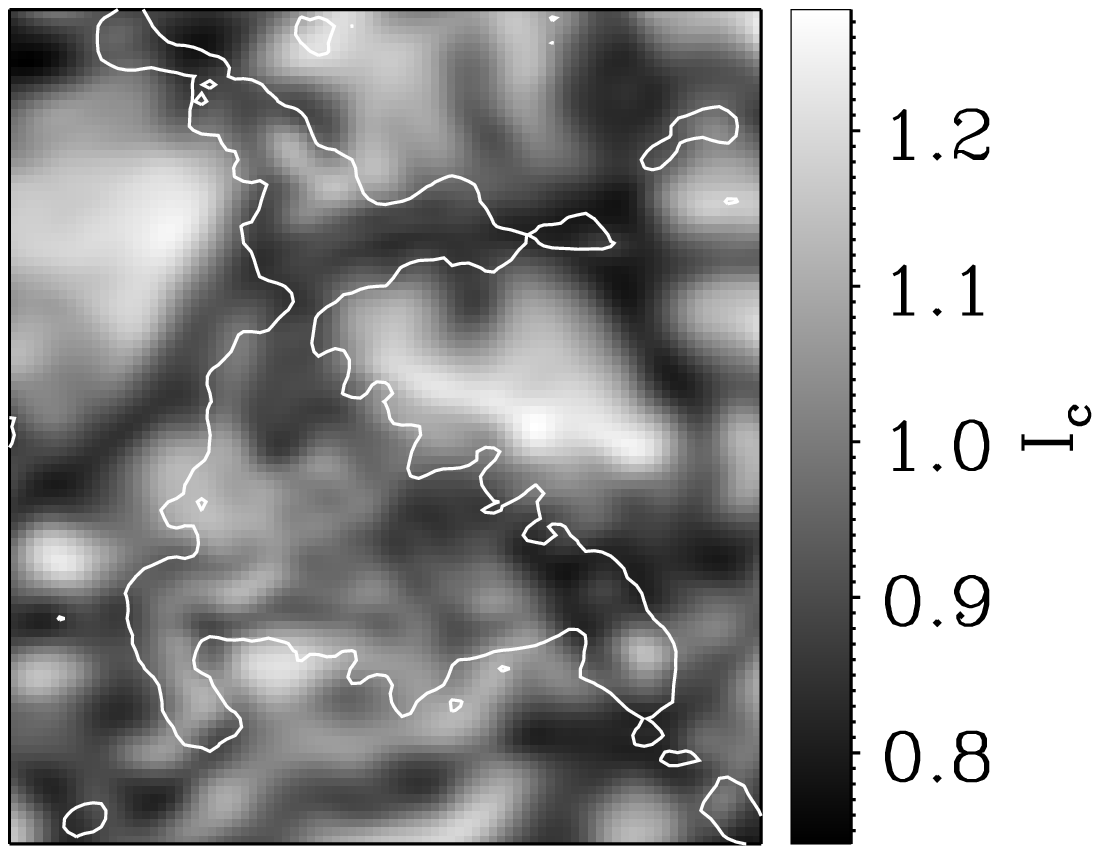}\\
\includegraphics[width=0.25\textwidth, bb=173 434 500 686]{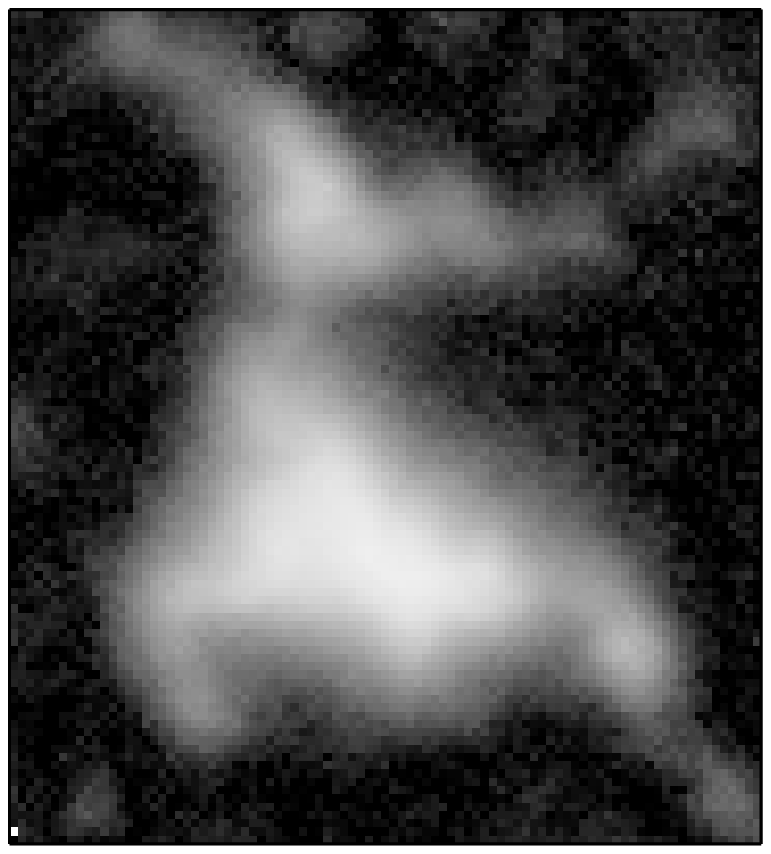}
\hspace{-1.5cm}
\includegraphics[width=0.25\textwidth, bb=173 434 500 686]{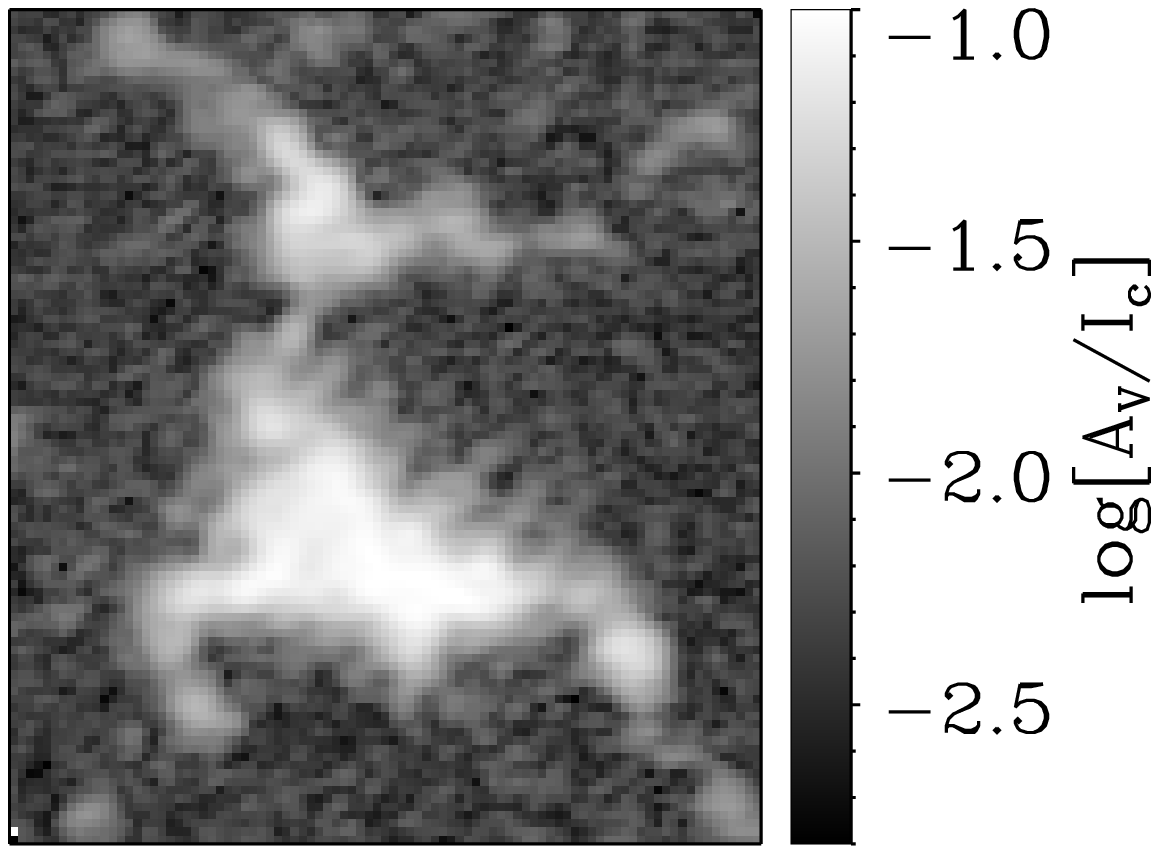}\\
\includegraphics[width=0.25\textwidth, bb=173 434 500 686]{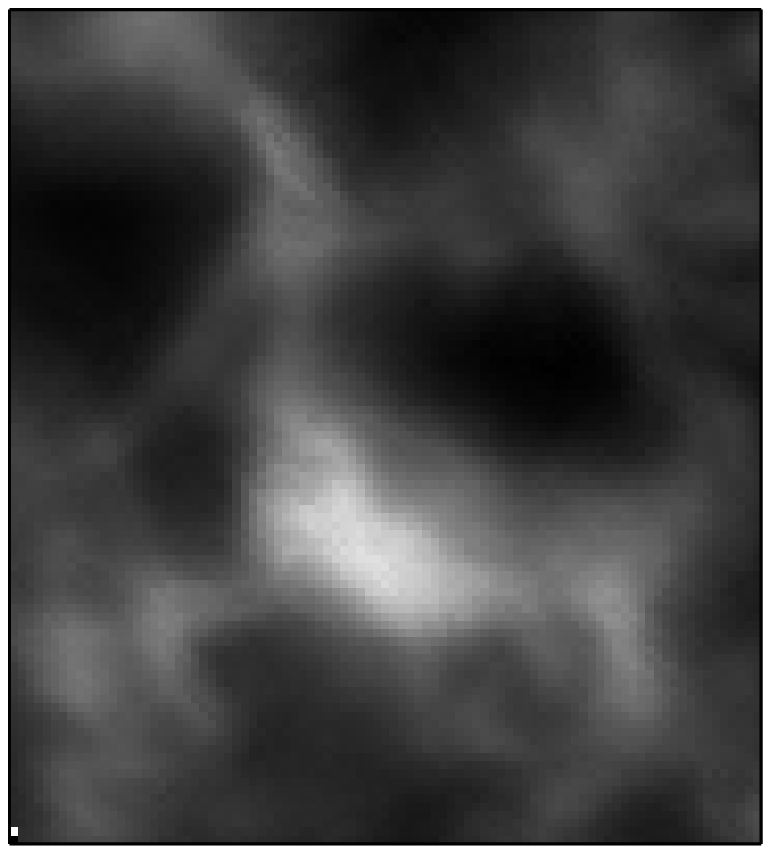}
\hspace{-1.5cm}
\includegraphics[width=0.25\textwidth, bb=173 434 500 686]{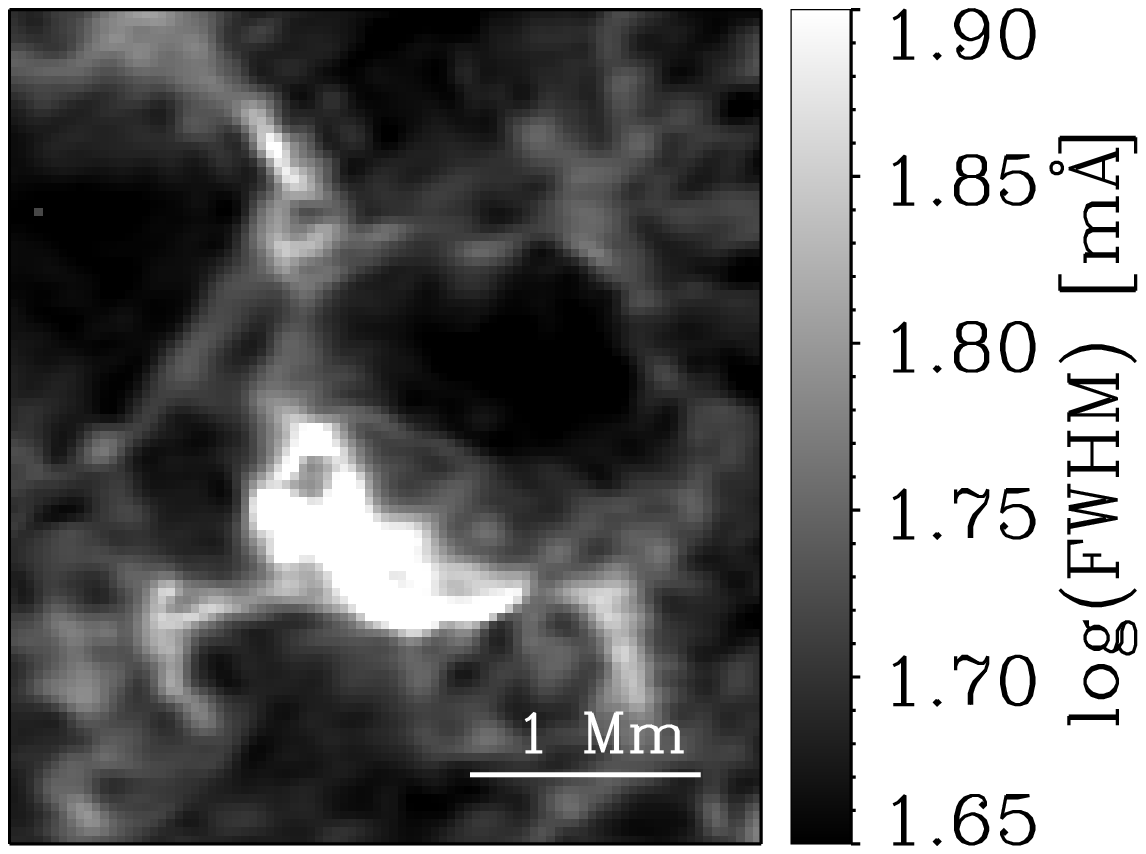}}
\caption{From top to bottom, continuum intensity, circular polarization amplitude and full width at half maximum of the intensity profile for the stongest flux region in the IMaX field of view. Left panels display the observables of the original data, while the right ones show the same scenes after phase-diversity reconstruction. The white contours represent the isocontours of log$[A_V/I_c]=-2.3,-1.9$ for the non recostructed and the reconstructed data, respectively. The symbol $A_V$ represents the amplitude of the Stokes $V$ profile.}
\label{mapas_region}
\end{figure}

\begin{eqnarray}
\delta A &=&\sigma \frac{\int_{\lambda_i}^{\lambda_f} V(\lambda)\mathrm{d}\lambda}{\int_{\lambda_i}^{\lambda_f} \mid V(\lambda)\mid \mathrm{d}\lambda} \\
\delta a &=& \frac{a_b-a_r}{a_b+a_r}
\label{eq}
\end{eqnarray}
The symbols $\lambda_i=5250.0164$ \AA\ and $\lambda_f=5250.4014$ \AA\ are the
initial and final wavelengths of the integral over the observed profile. The
symbol $\sigma$ is the sign of the Stokes $V$ blue lobe. The $a_b$ and $a_r$ symbols are
the unsigned blue and red lobe amplitudes, respectively. The non-linearity of
the equations to compute asymmetries make it difficult to evaluate the
relevance of instrument errors. The first error source to take into account is
the photon noise. We have verified (simulating Stokes $V$ profiles as derivatives of a gaussian that match IMaX intensity data and including gaussian noise) 
that the photon noise does not introduce a
bias in the calculation, only spreads the results, the relative errors in the
area and amplitude asymmetries being at most $\pm 15$\%. 

Calibrations performed at the launch station in Kiruna, proved that there
exist considerable amounts of crosstalk between linear and circular polarization
introduced by the ISLiD \citep{gandorfer_11}. The relative amounts of Stokes $Q$ and Stokes $U$
change with time due to image rotation and cannot be predicted for a specific data set.
When IMaX observes in a vectorial mode, all four Stokes parameters
are measured and multiplication by the inverse of the Mueller matrix takes
into account all crosstalk sources (leaving as unknown only the zero of
the azimuth angle). In the longitudinal observing mode used here \citep[L12-2, 
refer to][]{valentin_11} only Stokes $I$ and Stokes $V$ are measured making a 
proper correction of crosstalk impossible. However, we expect the impact of this
crosstalk to be negligible for the investigation performed in this work that 
concentrates on a relatively large network patch. First of all, the fields
in these regions are known to be predominantly vertical and all horizontal
fields observed with IMaX in other full vector maps show that the
linear polarization signals mainly corresponded to internetwork elements 
\citep{danilovic_10}. Second,  
inspection of similar network patches observed in the IMaX vector mode V5-6
\citep[bottom part of Figure 3 in][]{sami_10}, show that they never display
linear polarization signals above 3 sigma levels at the network concentration, neither 
inside it nor at the boundaries, where a possible canopy could 
generate some linear polarization. We thus conclude
that in the region under study, the known crosstalk in the longitudinal mode of
IMaX can be ignored even though it cannot be accounted for as is done
in the IMaX vector modes.

Finally, the effect of the finite spectral (and spatial) resolution, as well as
the effect of the phase-diversity reconstruction, is studied in
\cite{andres_11} for the special case of MHD simulations. These authors
conclude that the spatial degradation tends to understimate both amplitude and
area asymmetries. Degrading the spectral resolution, however, tends to
oveerstimate the area asymmetry while it does not affect strongly the amplitude
asymmetries, confirming earlier results of \cite{sami_stenflo_86}. After
applying the reconstruction algorithm, the computed asymmetries resemble the
original ones, i.e., for MHD simulations at full spectral and spatial
resolution. However, we must note that these conclusions depend on the actual
shape of the profiles. The work of \cite{sami_86} predicts (for the IMaX case)
an overstimate of the area asymmetries by a factor $\sim 1.6$ and an
understimate of the amplitude asymmetries by a factor $\sim 0.8$. Consequently,
strongly asymmetric/antisymmetric profiles observed by IMaX do correspond to
profiles with large/small asymmetries at full spectral resolution, although the
exact values may differ.

\begin{figure*}[!t]
\centering{
\includegraphics[width=0.33\textwidth, bb=127 406 472 700]{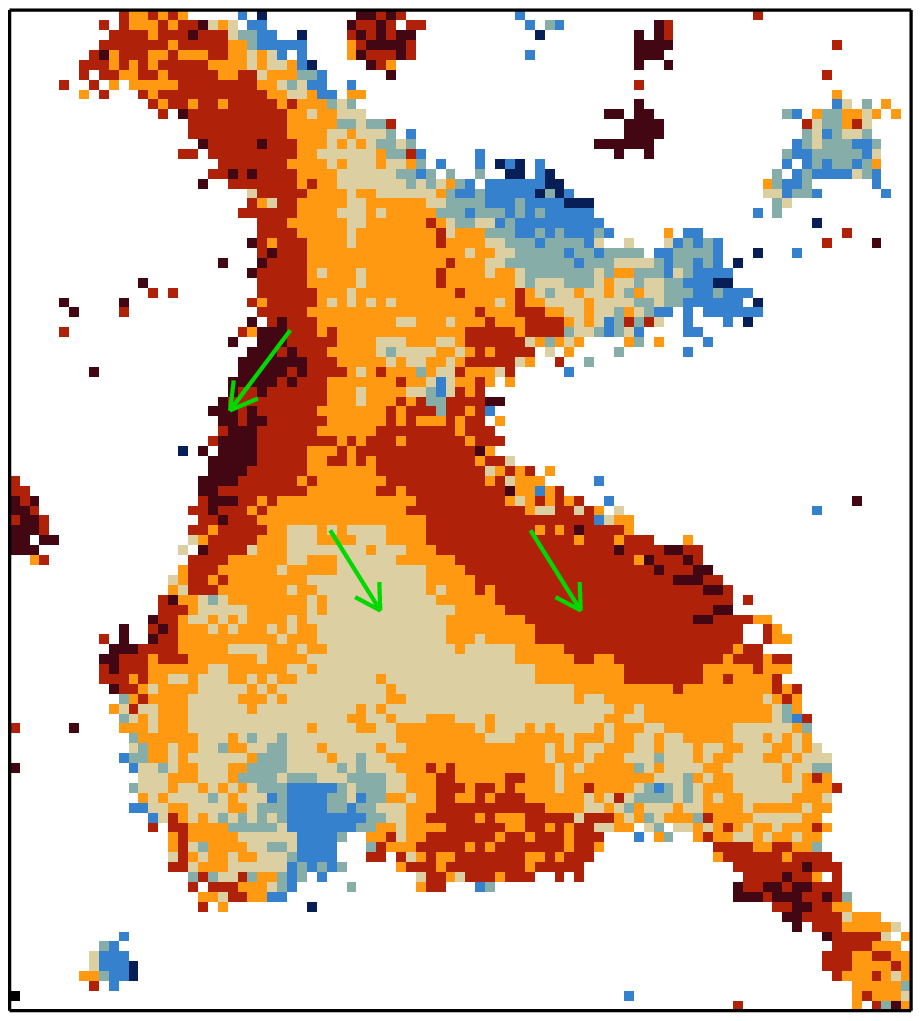}
\hspace{-1cm}
\includegraphics[width=0.33\textwidth, bb=127 406 472 700]{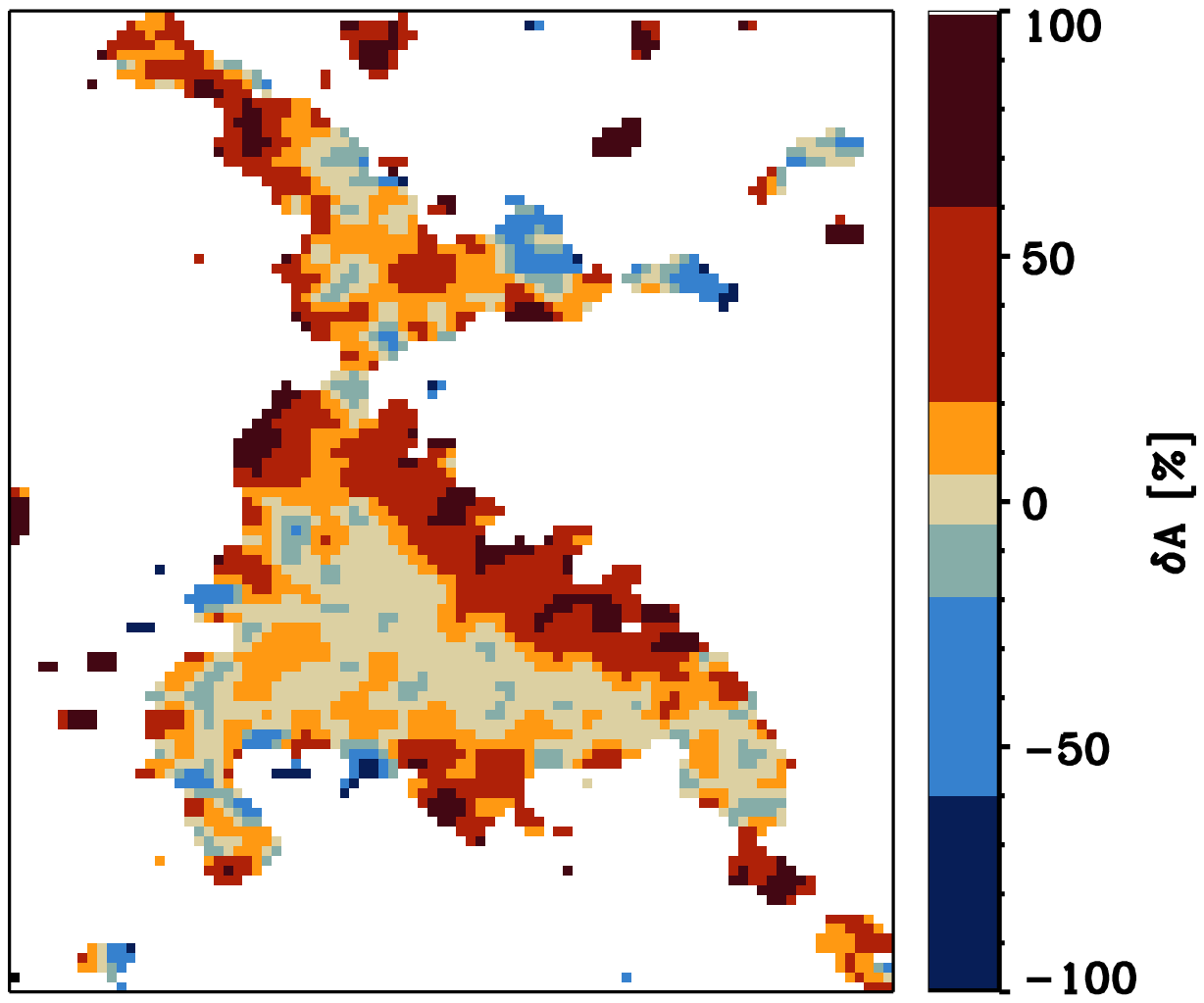}\\
\includegraphics[width=0.33\textwidth, bb=127 406 472 700]{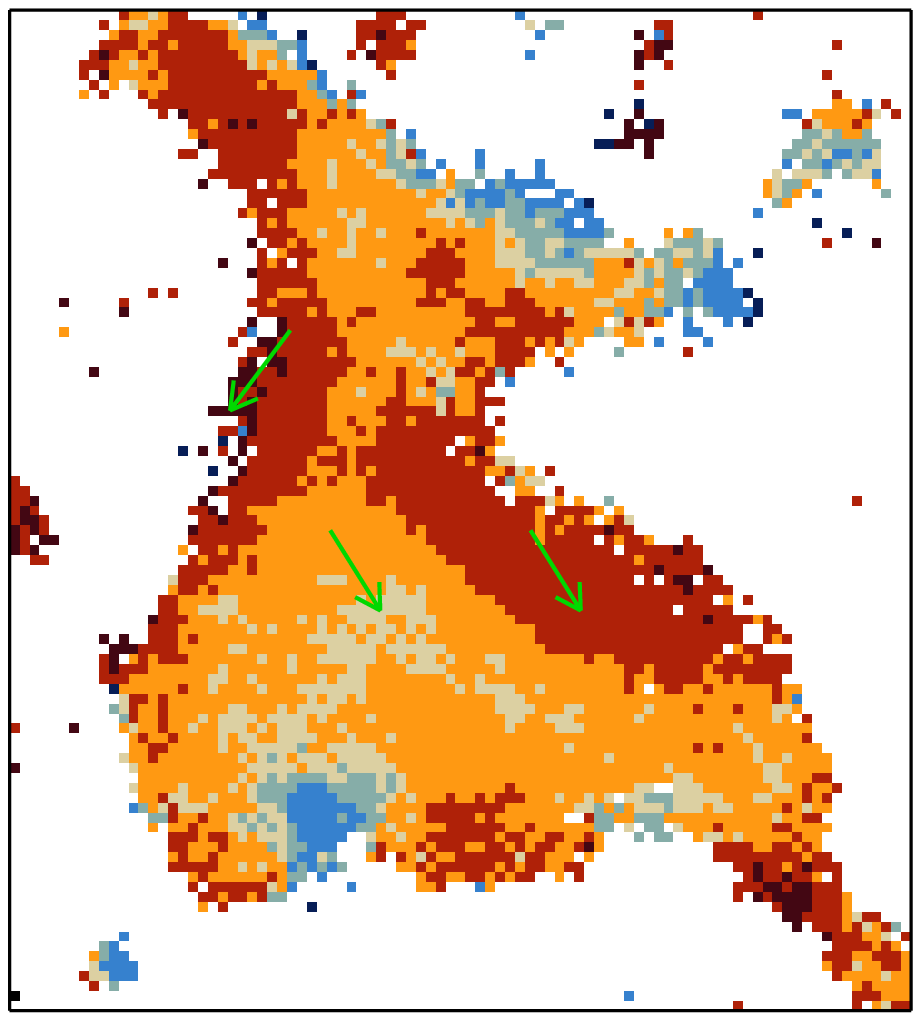}
\hspace{-1cm}
\includegraphics[width=0.33\textwidth, bb=127 406 472 700]{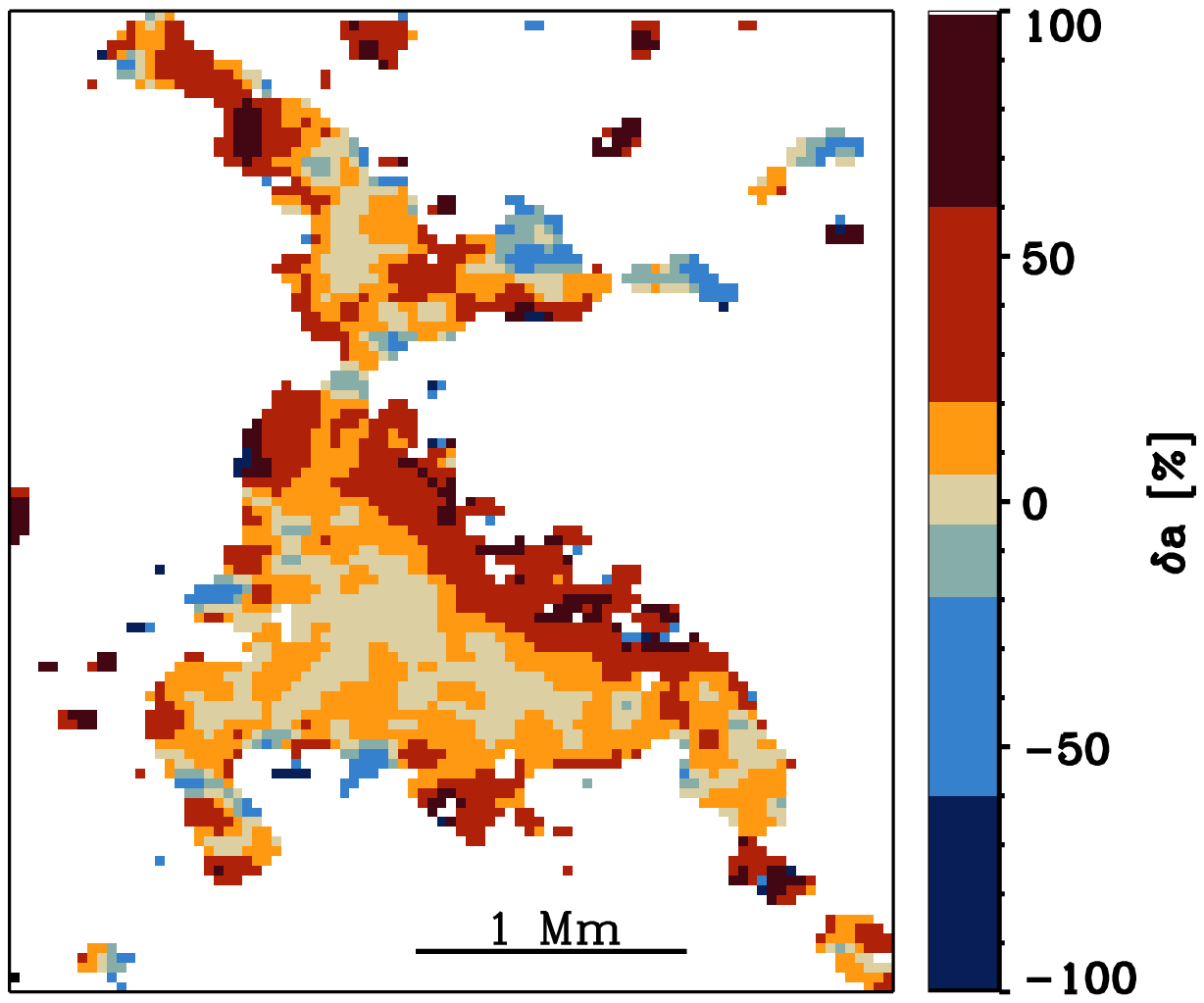}}
\caption{Top panels represent the area asymmetry of the circular polarization profiles of the selected strong flux patch. Bottom panels display the amplitude asymmetry. Right (left) panels present the results showing the reconstructed (non-reconstructed) data. The areas in white represent those pixels where the signal to noise ratio of the Stokes $V$ profiles is lower than 5. Grey arrows are the locations of the profiles displayed in Fig. \ref{perfiles}.}
\label{mapas_asim}
\end{figure*}

\section{Circular polarization asymmetries and the magnetic field geometry}

Figure \ref{mapas_asim} shows maps of the area and amplitude asymmetries of the observed Stokes $V$ profiles in the selected patch. The left panels correspond to the non-reconstructed data and the right  ones to the phase-diversity reconstructed ones. All four panels display the same pattern: the smaller asymmetries tend to be located in the inner parts of the structure, while the larger ones lie at the borders, close to where the signal drops below the noise level. The larger asymmetries situated at the borders of the structure have dominantly positive values, but negative values are also found (e.g. along the top edge of the feature). The spatial pattern for the two types of asymmetries is very similar, the correlation coefficient between amplitude and area asymmetries being 0.9.

\begin{figure}[!t]
\includegraphics[width=0.5\textwidth]{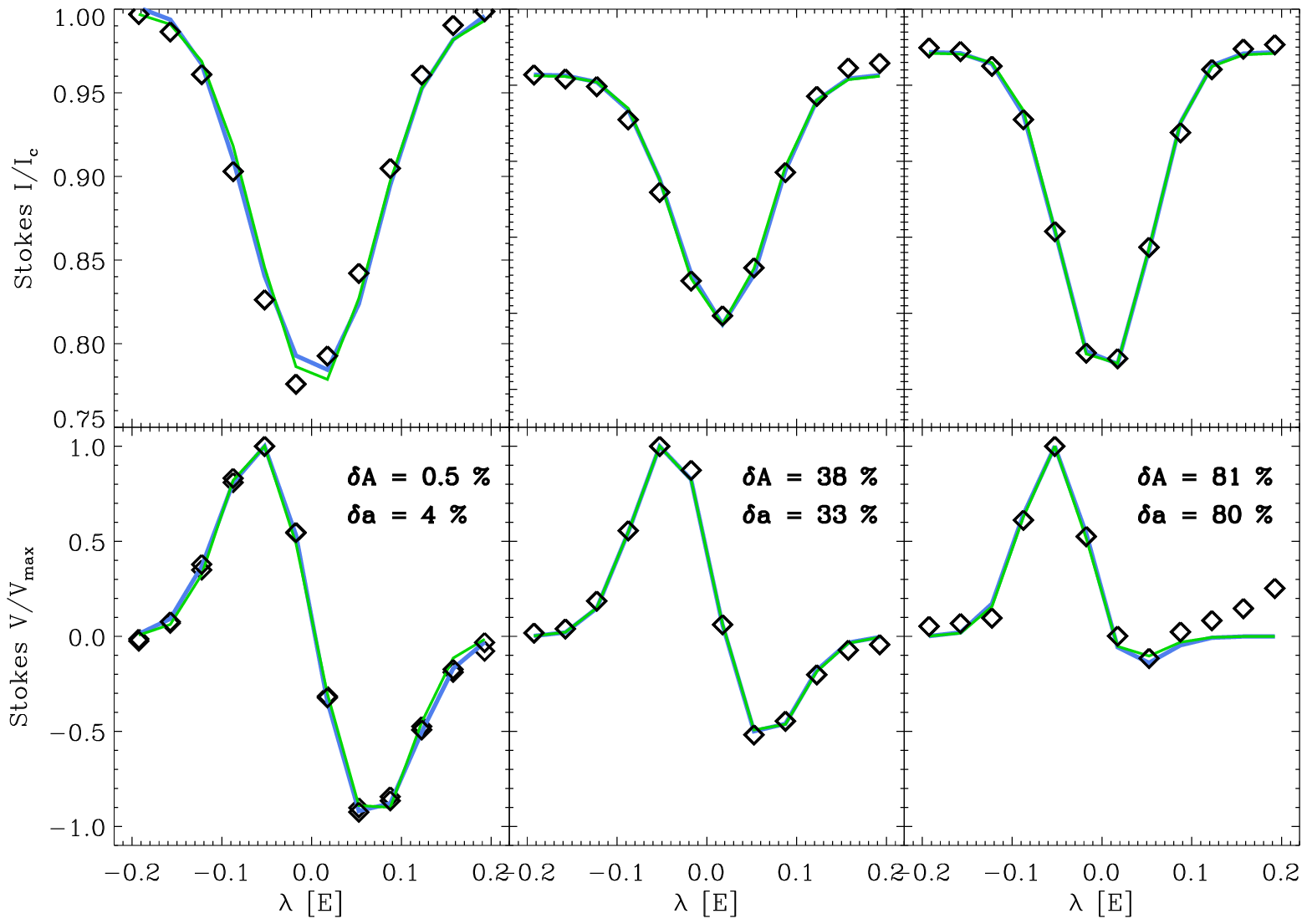}
\caption{Examples of Stokes $V$ profiles. The left panels display a respresentative of the relatively antisymmetric profiles found in the center of the structure, middle panels the more asymmetric ones found near the borders of the patch, while the right panels depict a one-lobed profile at the edge of the magnetic patch. The exact locations are marked in Fig. \ref{mapas_asim} by small grey arrows. The observed profiles are represented by the rhombs and the best fit using an inversion algorithm based on a discontinous model along the optical depth (see the text) is displayed with a blue line. The inversion with the inclination forced to zero is represented by the green lines and the inversion with the inclination as a free parameter is displayed with the blue lines.}
\label{perfiles}
\end{figure}

Figure \ref{perfiles} displays three sets of Stokes $I$ and $V$ profiles, extracted from the positions marked with small grey arrows in Fig. \ref{mapas_asim}. The Stokes $V$ profile displayed in the left panels ($\delta A=0.5$\% and $\delta a=4$\%) is representative of the nearly antisymmetric profiles found in the central part of the patch. The profile in the middle panels has larger area and amplitude asymmetries (38\% and 33\%, respectively), as found closer to the boundary of the magnetic feature. Finally, the profile in the rightmost panels is a one-lobed profile, with extreme asymmetries: 81\% area asymmetry, and 80\% amplitude asymmetry. Such profiles are found almost exclusively at the very edge of the feature.

In order to check if the expanding flux-tube model applies to our observations,
we use the SIRJUMP code, which is a modification of the SIRGAUS code
described in \cite{sirgaus}. The two codes follow the inversion
strategy of SIR \citep{basilio_92}. SIRJUMP has recently been
employed to interpret single-lobed Stokes $V$ profiles in the quiet Sun by \cite{alberto_12}.  
With SIRJUMP we retrieve the parameters of a model
atmosphere showing  discontinuous stratifications of some of the parameters
along the line of sight. The position of the jump (i.e., the discontinuity) is a free
parameter in the inversion scheme. The code can put it below optical
depth unity, which would indicate that no discontinuity is needed to explain the
observed profiles.  Above and below the jump, the magnetic field and the line-of-sight
velocity are assumed to be constant with optical depth.
The temperature stratification is modified with respect to an initial guess by the inversion algorithm in three nodes and
interpolated between them. The microturbulent velocity and
the inclination of the magnetic field vector to the line of sight
are height independent. Since we do not have the information of
the linear polarization, we cannot retrieve the  magnetic field azimuth, and can barely infer the inclination. 
Therefore, we perform two different inversions, one forcing the field to be aligned with the line-of-sight 
and another one leaving the inclination as a free parameter. The macroturbulent
velocity is fixed to 2.06 km s$^{-1}$, which matches the broadening
induced by the Point Spread Function of the IMaX instrument.

\begin{figure*}[!t]
\centering{
\includegraphics[width=0.28\textwidth, bb=80 50 260 340]{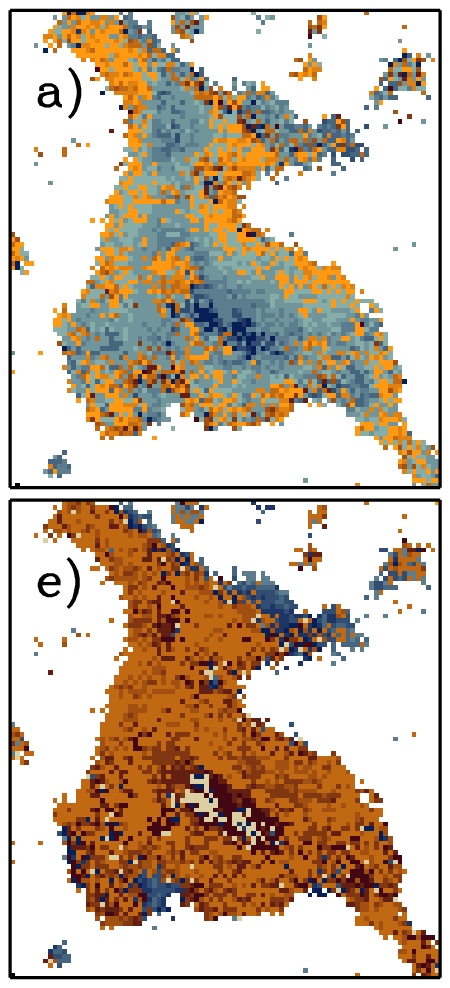}
\hspace{-1.65cm}
\includegraphics[width=0.28\textwidth, bb=80 50 260 340]{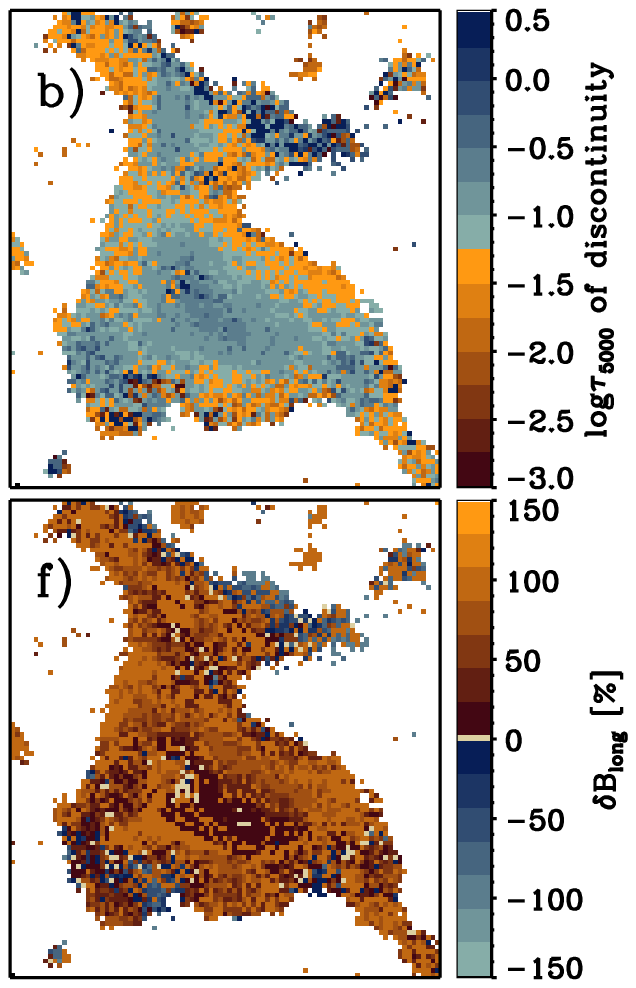}
\hspace{0.5cm}
\includegraphics[width=0.28\textwidth, bb=80 50 260 340]{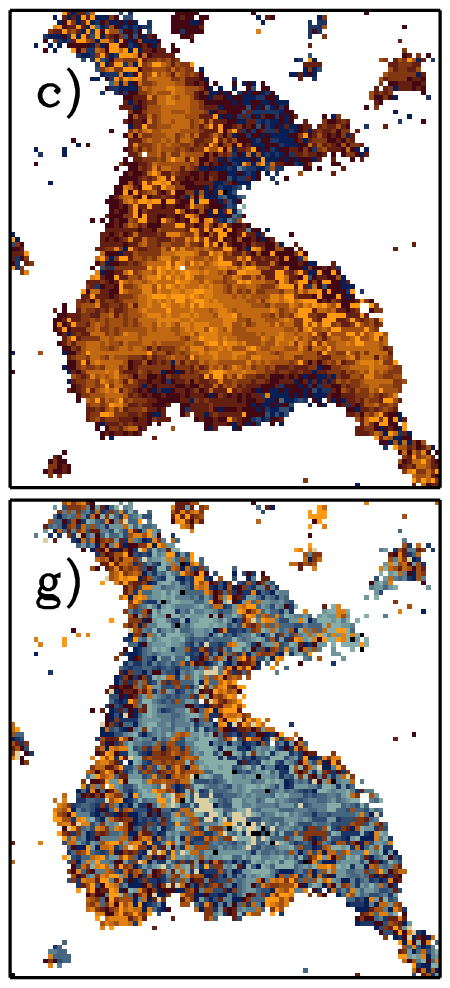}
\hspace{-1.65cm}
\includegraphics[width=0.28\textwidth, bb=80 50 260 340]{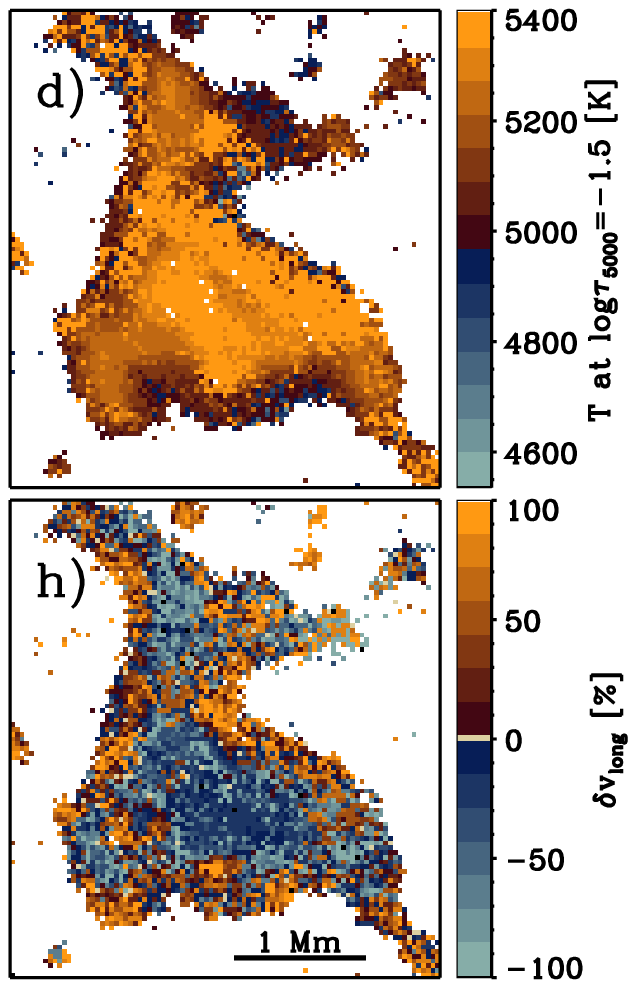}}
\caption{Results of the two inversions of the original data based on a model that has a discontinuity along the optical axis. Panels a, c, e, and g display the inferred quantities for the inversion in which the inclination of the magnetic field is a free parameter. Panels b, d, f, and h represent the retrieved parameters from the inversion in which we have fixed the inclination of the magnetic field with respect to the line-of-sight to 0$^\circ$. Similar results can be derived from the phase diversity reconstructed data. Panels a-d show the optical depth of the discontinuity (left) and the temperature at log${\tau_{5000}}=-1.5$ (right). Panels e-h display the relative differences of the line-of-sight magnetic field (left) and velocity (right) above and below the discontinuity (see Eq. 3)}
\label{mapas_inv}
\end{figure*}

Figure \ref{mapas_inv} displays the results of the two different inversions. Top left panels show the optical depth at which the discontinuity occurs for the inversion with the inclination as a free parameter (panel 4a) and the inversion forcing the inclination to 0$^\circ$ (panel 4b). In both cases, the results are very similar, i.e., they do not depend much on the assumed geometry of the magnetic field. For log$\tau_{5000}  \gtrsim 0.5$ the contribution to the spectral line is small and the discontinuity is not well constrained. In spite of the noise some trends can be seen. In the central part of the structure we retrieve a magnetic field with no clear discontinuity. As we move from the center towards the edges of the structure, the discontinuity appears at increasingly higher layers. Therefore, the inversions made pixel by pixel have naturally retrieved the expansion of a magnetic flux concetration with height in the solar atmosphere. At all heights (below and above the discontinuity), the pattern of the velocity is qualitatively similar, following the local distorted granulation.

In the lower panels of the same figure we display the relative differences of the longitudinal components of the magnetic field (panels e and f) and the velocity (panels g and h) above and below the discontinuity. Note that, since we do not have linear polarization, the only trustable quantity is the longitudinal component of the magnetic field. We have used the following definitions:

\begin{eqnarray}
\delta B_{long} &=& \frac{|B_{long}^{a}| - |B_{long}^{b}|}{|B_{long}^{a}| + |B_{long}^{b}|} \nonumber \\
\delta v_{long} &=& \frac{|v_{long}^{a}| - |v_{long}^{b}|}{|v_{long}^{a}| + |v_{long}^{b}|},
 \end{eqnarray}
where the superindex $a$ and $b$ refer to the quantities above and below the discontinuity, respectively. 
From these figures, we can see that both the longitudinal magnetic field and the line-of-sight velocity 
are almost constant with optical depth at the core of the structure, coinciding with strong kG fields. As we move towards the borders of the structure (where the longitudinal component of the magnetic field is of the order of 100 G), the relative variation of both quantities increases. Note that the gradient of the longitudinal magnetic field changes sign at the regions where the area and amplitude asymmetries were negative. 

\section{Conclusions}

We have computed the area and amplitude asymmetries of the Stokes $V$ profiles of the Sunrise/IMaX dataset with the best spectral sampling (35 m\AA). We selected the strongest magnetic flux accumulation, which belongs to the network pattern, in the IMaX FOV. The core of the structure is, on average, characterized by an area asymmetry of 0.2\% and an amplitude asymmetry of 4.4\% (with an upper limit for the error of $\pm 1.3$\% for both averaged area and amplitude asymmetries). In the outermost parts of the network patch, the Stokes $V$ profiles have the largest area and amplitude asymmetries, even aproaching 100\% (one-lobed profiles). The fact that we are seeing a spatial variation 
of Stokes $V$ asymmetries means that, for the first time, we are resolving the internal structure of a magnetic patch belonging to the photospheric network. The non negligible amplitude asymmetry inside the magnetic feature suggests that unresolved motions are still present (definitely gradients along the line of sight and probably horizontal gradients). 

We inverted our data using the SIRJUMP code, which models the gradients of magnetic field and line-of-sight velocity with an abrupt discontinuity at a certain optical depth. From two different assumptions on the inclination of the field, we inferred that the observed network patch is made of resolved kG fields which is consistent with the work of \cite{lagg_10}. Moreover, the inversions naturally retrieve a geometry corresponding to a magnetic field expanding with height. 

The spatial distribution of the Stokes $V$ asymmetries and the inversions applied to these data suggest a geometry similar to the classical model of magnetic features expanding with height \citep{spruit_76, sami_93} that takes pressure balance and magnetic flux conservation with height into account and is consistent with three-dimensional magneto-hydrodynamical simulations \citep{lotfi_09}. In such a model, the central core of the structure traces the body of the flux tube, containing the vertical kG fields, while the external parts correspond to the canopy, i.e. expanding, inclined field overlying the quiet photosphere (this expansion is caused by the need to conserve magnetic flux for decreasing gas pressure with height). 

The standard model for the creation of asymmetries in the presence of such a canopy is based on the presence 
of a discontinuity in both the magnetic field strength and the line-of-sight velocity at the base of the canopy (i.e. abrupt gradients in such quantities along the optical depth). In the past, it was assumed that there is no velocity flow above the canopy's base but there is a downflow below it, with the magnetic field being present only above the canopy base. This geometry produced a positive area (and amplitude) asymmetry \citep{grossmann_88, sami_89}, with nearly one-lobed profiles in the outer parts of the flux tubes \citep{sami_89} in agreement with the observations. 

Our results for the asymmetries (see Fig. \ref{mapas_asim}) are in 
agreement with this general picture. However, there are some locations along the edge of our feature where the area (and amplitude) asymmetry is negative, indicating that either the magnetic field or the velocity field gradient must change sign there. We know that the kG network fields are embedded in the more disorganised weakly magnetized quiet photosphere. Hence, we would expect a smooth transition between the organized kG fields to the 
tangled weak photospheric fields (note that we have selected the profiles above the noise level, i.e., the borders of the structure depend on the amplitude recquired for the profiles). If those regions are mostly tracing the quiet photosphere (note that the height of the discontinuity is located at lower layers and that the temperature is colder than that of the central part of the structure), it is not surprinsing to find gradients of magnetic field and velocity that change sign as compared to the classical canopy overlying a non-magnetized photosphere. 

The observations presented in this letter support the interpretation of the network as kG fields expanding with height. In contrast to what was assumed in the past, we find velocity patterns above and below the canopy that follow the local distorted granulation. To strengthen the model for strong flux concentrations as the one studied in this letter, in the future we need to include in our analysis the constraints introduced by the linear polarization and, more importantly, the simultaneous information of many spectral lines to trace the canopy at different heights.

\begin{acknowledgements}

We specially thank Manuel Collados Vera for very helpful discussions and for carefully reading the manuscript. The Spanish contribution has been funded by the Spanish Ministry of Science and Innovation under projects ESP2006-13030-C06, AYA2009-14105-C06 (including European FEDER funds), and  AYA2010-18029 (Solar Magnetism and Astrophysical Spectropolarimetry). The German contribution to Sunrise is funded by the Bundesministerium f\"ur Wirtschaft und Technologie through Deutsches Zentrum f\"ur Luft- und Raumfahrt e.V. (DLR), Grant No. 50 OU 0401, and by the Innovation fond of the President of the Max Planck Society (MPG). This work has been partially supported by the WCU grant (No. R31-10016) funded by the Korean Ministry of Education, Science and Technology.

\end{acknowledgements}


\end{document}